\documentclass{sig-alternate}

\usepackage{cite}
\usepackage{graphicx}
\usepackage{amsmath}
\usepackage{algorithmic}
\usepackage{url}
\usepackage{float}
\usepackage{xcolor}

\usepackage{booktabs}
\usepackage{multirow}

\usepackage{caption}
\usepackage{subcaption}

\newcommand{\aspas}[1]{{``#1''}}

\newcommand{\sbar}[2]{{\color{darkgray}\rule{\dimexpr 1cm * #1 / #2}{6pt}}}
\newcommand{\sbarn}[2]{{\color{lightgray}\rule{\dimexpr 1cm * #1 / #2}{5pt}}}

\begin{document}

\setcopyright{acmcopyright}





\title{Predicting the Popularity of GitHub Repositories}

\numberofauthors{1}

\author{
  \alignauthor{Hudson Borges, Andre Hora, Marco Tulio Valente}\\
  \affaddr{ASERG Group, Department of Computer Science (DCC)}\\
  \affaddr{Federal University of Minas Gerais (UFMG), Brazil}\\
  \email{\{hsborges, hora, mtov\}@dcc.ufmg.br}
}


\maketitle

\begin{abstract}
GitHub is the largest source code repository in the world. It provides a git-based source code management platform and also many features inspired by social networks. For example, GitHub users can show appreciation to projects by adding {\em stars} to them.
Therefore, the number of stars of a repository is a direct measure of its popularity. In this paper, we use multiple linear regressions to predict the number of stars of GitHub repositories. These predictions are useful both to repository owners and clients, who usually want to know how their projects are performing in a competitive open source development market. In a large-scale analysis, we show that the proposed models start to provide accurate predictions after being trained with the number of stars received in the last six months. Furthermore, specific models---generated using data from repositories that share the same growth trends---are recommended for repositories with slow growth and/or for repositories with less stars.
Finally, we evaluate the ability to predict not the number of stars of a repository but its rank among the GitHub repositories. We found a very strong correlation between predicted and real rankings (Spearman's $\mathit{rho}$ greater than 0.95).
\end{abstract}


\section{Introduction}
\label{sec:introduction}

GitHub is the largest software repository in the world.
Despite git-based source code management services (init, clone, add, commit, push, etc), GitHub also supports social coding features. For example, developers can show appreciation to a repository by using the {\em star} button, which essentially plays the same role as the {\em like} button in other social networks. Therefore, the number of stars of a repository works like an easily accessible and reliable proxy to its popularity~\cite{Borges2015, Hudson2016}. In fact, the top-starred repositories in GitHub are widely known software projects, e.g., {\sc jquery/jquery} (a library to HTML scripting), {\sc torvalds/linux} (the Linux kernel), {\sc rails/rails} (a web framework for Ruby), and {\sc do\-cker/do\-cker} (an application container engine).

Due to the relevance of GitHub in modern open source development, researchers started to study the popularity of GitHub repositories. For example, a study by Zho et al.~shows that the adoption of standard folders (e.g.,~doc, test, examples) may have an impact on project code popularity~\cite{Zhu2014}. In another study, Aggarwal et al.\cite{Aggarwal2014} show that  popular projects tend to attract more documentation collaborators. Weber and Luo
attempted to differentiate popular and unpopular Python projects on GitHub using machine
learning techniques~\cite{Weber2014}.
Recently, we investigated the factors that impact the popularity of GitHub repositories, including programming language, application domain, repository owner (user or organization), age, and release frequency~\cite{Hudson2016}.  For this purpose, we collected historical data about the number of stars of 2,500 popular repositories. We also used this dataset to identify four patterns of popularity growth, which we called slow, moderate, fast, and viral.

In this paper, we extend our previous work by investigating the use of multiple linear regressions to {\em predict the popularity} of GitHub repositories. Prediction models have been successfully used to infer the popularity of content in other social networks, such as the number of views of YouTube videos~\cite{Roy2013, Pinto2013, Figueiredo2013} and the number of tweets associated to a given hashtag~\cite{tsur2012, ma2012, ma2013}. However, to our knowledge, we are the first to attempt to predict the popularity---measured by the number of stars---of software projects hosted at GitHub. Specifically, we compute and investigate multiple linear regression models over two types of data: {\em generic} and {\em specific}. By generic, we refer to models produced from the complete dataset considered in this paper, which includes historical data about the number of stars of 4,248 popular GitHub repositories. By specific, we refer to models produced from repositories that share similar growth trends. These trends are inferred using the KSC algorithm~\cite{Yang2011}, which clusters time series with similar shapes.

We address three major research questions in the paper:

\begin{itemize}

\item {\em RQ \#1: What is the accuracy of the generic prediction models?} We report the Relative Squared Error (RSE) of the regression models computed using the time series of number of stars of all projects in our dataset.

\item  {\em RQ \#2: What is the accuracy of the specific prediction models?} First, using the KSC clustering algorithm, we identify four major growth trends among the systems in our dataset. Then, we evaluate the accuracy of the regression models computed over the time series of each cluster.

\item  {\em RQ \#3: What is the accuracy of the repositories rank as predicted using the generic and specific models?} In the previous RQs, our goal is to the predict the total number of stars, using generic and specific models. By contrast, in this final RQ, we evaluate the ability of both models to predict not the number of stars of a repository after a time, but its rank among the repositories in our dataset.

\end{itemize}

The remainder of this paper is organized as follows. First, we present the dataset used in the paper (Section~\ref{sec:dataset}) and discuss the methodology followed in the study (Section~\ref{sec:study-design}). Section~\ref{sec:results} presents our results, by exploring and discussing answers for the three proposed research questions. Finally, Section~\ref{sec:validity} discusses threats to validity and Section~\ref{sec:conclusion} concludes the paper.

\section{Dataset}
\label{sec:dataset}

The initial dataset used in this paper includes historical data about the top-5,000 public repositories with more stars in GitHub.
All data was obtained using the GitHub API, which provides services to search public repositories and to retrieve specific information about them (e.g., stars).
First, we collect basic data about the repositories (i.e., owner, stars, creation date, programming language, etc.).
Next, for each repository, we collect historical data about the number of stars.
For this purpose, we used a service from the API that returns all star events of a given repository.
For each star, these events store the date and the user responsible to starring the repository. However, GitHub API returns at most 100 events by request (i.e., a page) and at most 400 pages.
For this reason, it is not currently possible to retrieve all star events of systems with more than 40K stars, which is
the case of seven repositories: {\sc FreeCodeCamp} (112,397 stars), {\sc twbs/bootstrap} (95,293 stars), {\sc vhf/free-progra\-mming-books} (54,208 stars), {\sc mbostock/d3} (49,173 stars), {\sc angular/angular.js} (48,787 stars), {\sc FortAwesome/Font-Awesome} (41,621 stars),  {\sc facebook/react} (41,037 stars).
Moreover, 278 repositories have no main programming language identified. These repositories do not store source code, e.g., {\sc jlevy/the-art-of-command-line} (26,298 stars) or are moved/removed repositories, e.g., {\sc nodejs/no\-de-v0.x-archive} (37,354 stars).
Therefore, we also remove these repositories from the dataset.
Additionally, we only consider the stars gained in the last 52 weeks of each repository.
Thus, repositories with less than 52 weeks are also removed from the dataset (468 repositories).

\begin{figure}[!h]
  \centering
  \includegraphics[width=0.65\columnwidth, trim={0 1.5em 0 5em}, clip]{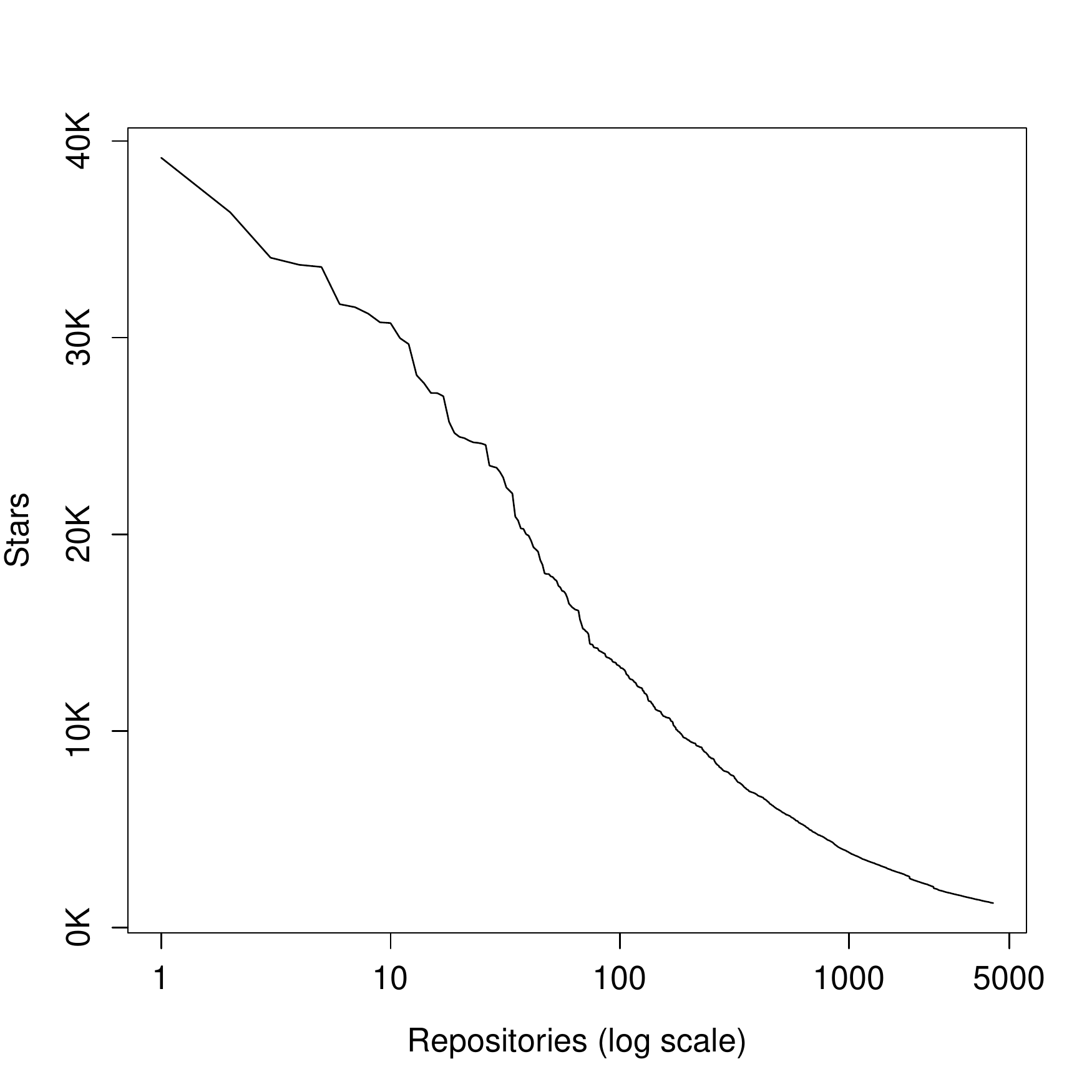}
  \caption{Repositories popularity}
\label{fig:dataset-stars}
\end{figure}

Figure~\ref{fig:dataset-stars} shows the number of stars of the 4,248 repositories in our dataset.
This number ranges from 39,149 stars ({\sc jquery/ jquery}) to 1,248 stars ({\sc mikeflynn/egg.js}).
As presented, the distribution is right skewed (quantiles 5\% = 1,307 stars and 95\% = 9,360 stars).
The mean and median number of stars are 3,393 and 2,240, respectively.
Table~\ref{tab:dataset-top} lists the top-10 repositories with more stars.
These repositories have at least 30K stars and belong to four different domains (Web Frameworks and Libraries, Software Tools, Documentation, and System Software).

\begin{table}[!h]
  \centering
  \caption{Top-10 repositories with more stars}
\label{tab:dataset-top}
  \begin{tabular}{@{}lcc@{}}
    \toprule
    \multicolumn{1}{c}{\textbf{Repository}} & \multicolumn{1}{c}{\textbf{Domain}} & \multicolumn{1}{c}{\textbf{\# Stars}}\\
    \midrule
    {\sc jquery/jquery}          & Web             & 39,149\\
    {\sc robbyrussell/oh-my-zsh} & Tools  & 36,373\\
    {\sc airbnb/javascript}      & Doc   & 34,064\\
    {\sc h5bp/html5-boilerplate} & Web             & 33,704\\
    {\sc meteor/meteor}          & Web             & 33,594\\
    {\sc torvalds/linux}         & System & 31,702\\
    {\sc daneden/animate.css}    & Web             & 31,549\\
    {\sc facebook/react-native}  & Web             & 31,217\\
    {\sc rails/rails}            & Web             & 30,779\\
    {\sc docker/docker}          & System & 30,742\\
    \bottomrule
  \end{tabular}
\end{table}

\begin{figure}[!h]
  \centering
  \includegraphics[width=0.65\columnwidth, trim={0 1.5em 0 5em}, clip]{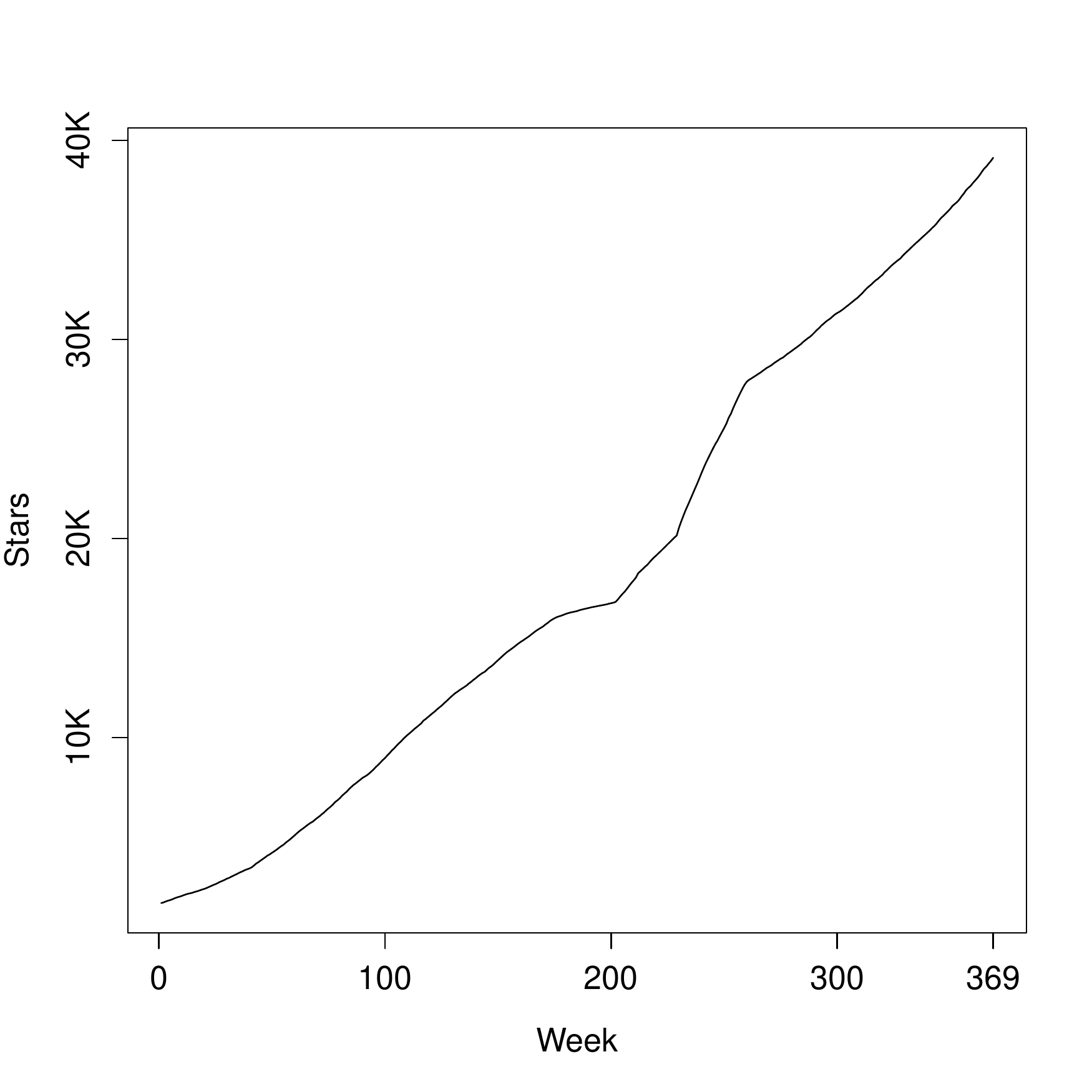}
  \caption{{\sc jquery/jquery} time series (369 weeks)}
\label{fig:dataset-jquery-timeseries}
\end{figure}

Next, we built the stars time series of each repository from the stars events.
These time series consist of the number of stars gained by week since the repository creation date up to April 25, 2016, when we collected our data.
As an example, Figure~\ref{fig:dataset-jquery-timeseries} shows the time series retrieved for {\sc jquery/jquery}, the most starred repository in our dataset.
This repository has 369 weeks (x-axis) and the number of stars increased from 1,692 stars to 39,149 stars (y-axis).

\section{Study Design}
\label{sec:study-design}

In this section, we detail the techniques and models used to predict the number of stars of GitHub repositories.
We also discuss how we evaluate the accuracy of these models.

\vspace{1em}\noindent{\bf Prediction Technique:}
We rely on multiple linear regression to predict the popularity of GitHub repositories.
Multiple linear regression differs from simple regression by considering that all variables are not equally important~\cite{freedman2009statistical}.
The general form of a multiple linear regression is as follows:
$$ Y_{t} = b_{0} + b_{1}X_{t_{1}} + b_{2}X_{t_{2}} + ... + b_{r}X_{t_{r}} $$
where $Y_{t}$ is the dependent variable (number of stars at week $t$), $X_{t_{i}}$ are the independent variables (stars in weeks $i$, $1 \leq i \leq r \leq t$), and $b_{j}$ are the regression coefficients ($0 \leq j \leq r \leq t$).

\vspace{1em}\noindent{\bf Estimating the Errors:}
To evaluate the accuracy of the models, we use the Relative Squared Error (RSE).
Assume that $N(r, t)$ is the real number of stars of a repository $r$ in the week $t$.
Moreover, assume that $\widehat{N}(r, t_{r}, t)$ is the number of stars \emph{predicted} for $r$ at the week $t$ from the popularity data of the first $t_{r}$ weeks.
The RSE for this prediction is given by\cite{Pinto2013}:
$$ RSE = \left ( \frac{\widehat{N}(r, t_{r}, t)}{N(r, t)} - 1 \right )^{2} $$

For a collection $\mathcal{R}$ of repositories, the mean Relative Squared Error (mRSE) is defined as the arithmetic mean of the RSE values of all repositories in $\mathcal{R}$, as given by:
$$ mRSE =  \frac{1}{\left |\mathcal{R}  \right |} * \sum_{r \in \mathcal{R}} \left ( \frac{\widehat{N}(r, t_{r}, t)}{N(r, t)} - 1 \right )^{2} $$

\begin{figure}[!h]
  \centering
  \includegraphics[width=0.65\columnwidth, trim={8em 38em 10em 10em}, clip]{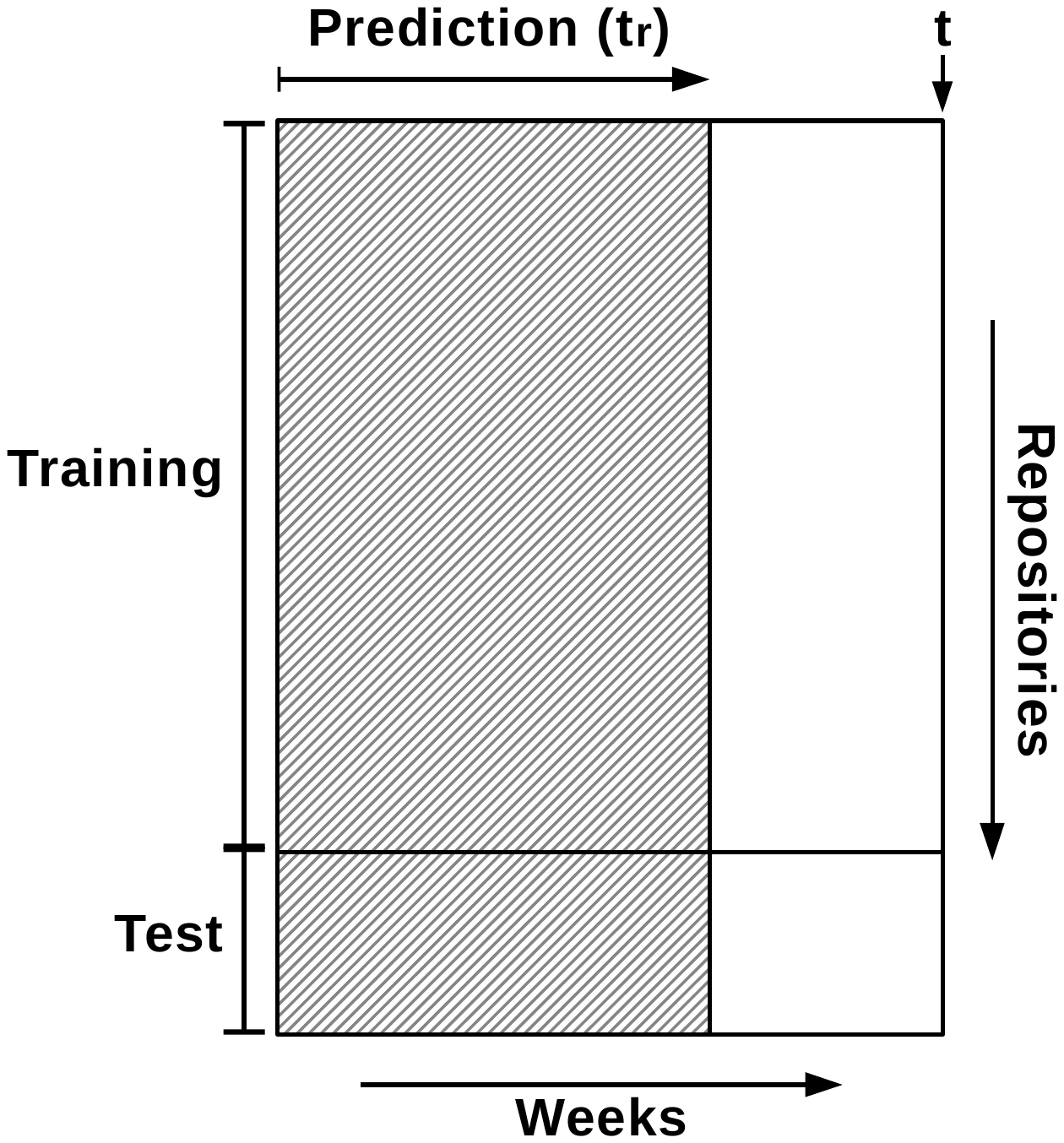}
  \caption{Cross Validation}
\label{fig:cross-validation}
\end{figure}

\vspace{1em}\noindent{\bf Cross-Validation:}
As ilustrated in Figure~\ref{fig:cross-validation}, we perform cross-validation to assess the prediction models.
We use 10 folds, i.e., the repositories are randomly partitioned in 10-folds and we use nine folds to build the prediction models (training set) and the remaining fold to evaluate their accuracy (validation set).

\begin{figure}[!h]
  \centering
  \includegraphics[width=0.65\columnwidth, trim={0 2em 0 5em}, clip]{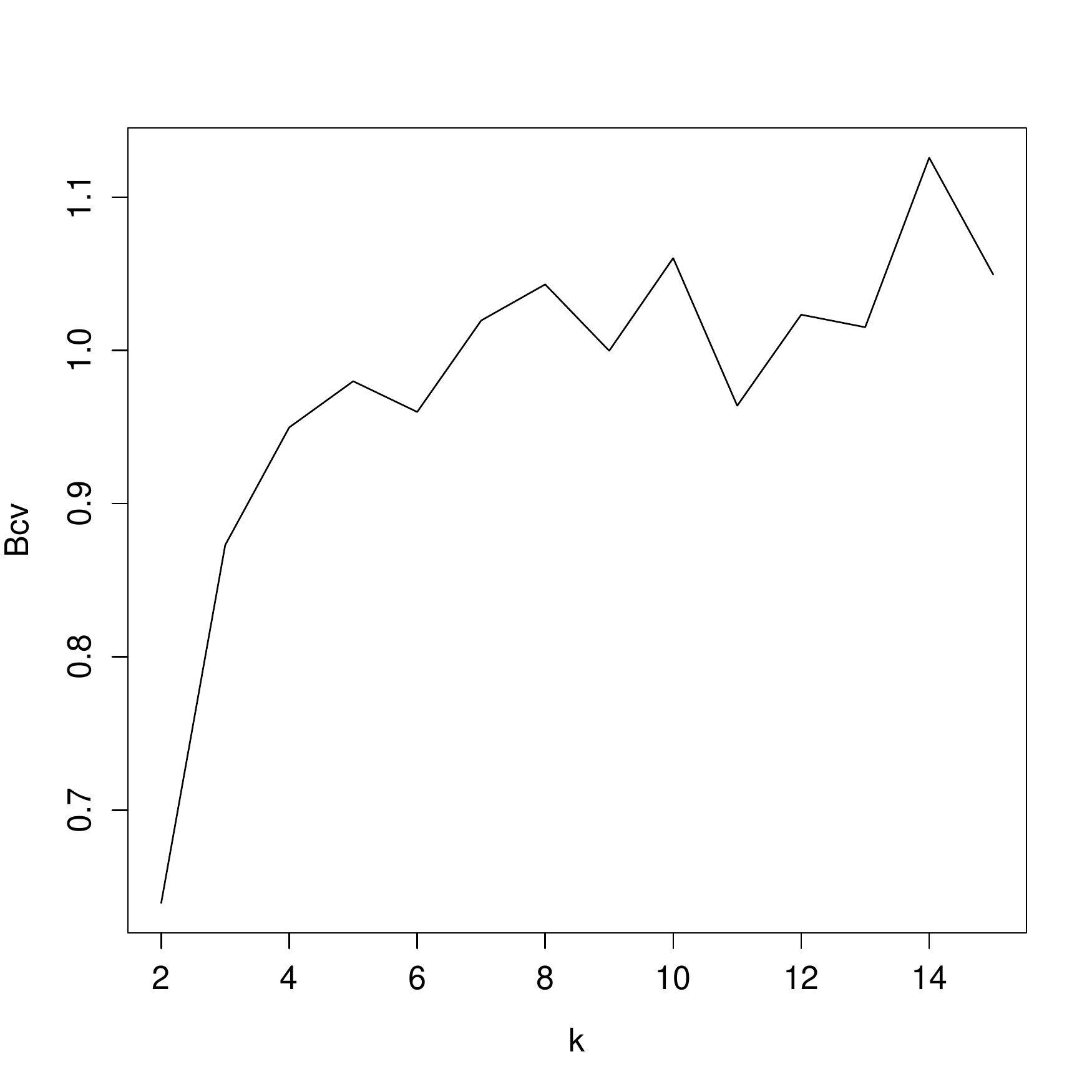}
  \caption{$\beta_{CV}$ ($2 \leq k \leq 15)$}
\label{fig:bcv}
\end{figure}

\vspace{1em}\noindent{\bf Generic and Specific Models:}
We generate models for two datasets: {\em generic} and {\em specific}.
By generic, we refer to models produced from the complete dataset, i.e.,~from the time series with the number of stars collected for 4,248 repositories. By specific, we refer to models produced from repositories that shared similar growth trends.
As in our previous work~\cite{Hudson2016}, we rely on the KSC algorithm~\cite{Yang2011}  to identify growth trends in our dataset.
This algorithm clusters time series with similar shapes using a metric that is invariant to scaling and shifting.
In other words, each cluster groups time series that share similar growth trends. Particularly, to answer RQ \#2 we produce specific models considering only the time series in each cluster.
We use the $\beta_{CV}$ heuristic~\cite{Menasce2001} to define the best number $k$ of clusters.
$\beta_{CV}$ is defined as the ratio of the coefficient of variation of the intracluster distances and the coefficient of variation of the intercluster distances.
The smallest value of $k$ after which the $\beta_{CV}$ ratio remains roughly stable should be selected.
In our dataset, the values of $\beta_{CV}$ stabilize for $k = 5$ (see Figure~\ref{fig:bcv}).
Therefore, we configure KSC to produce five clusters.

\begin{figure}[!h]
  \centering
  \includegraphics[width=1\linewidth, trim={0 0em 0 0em}, clip]{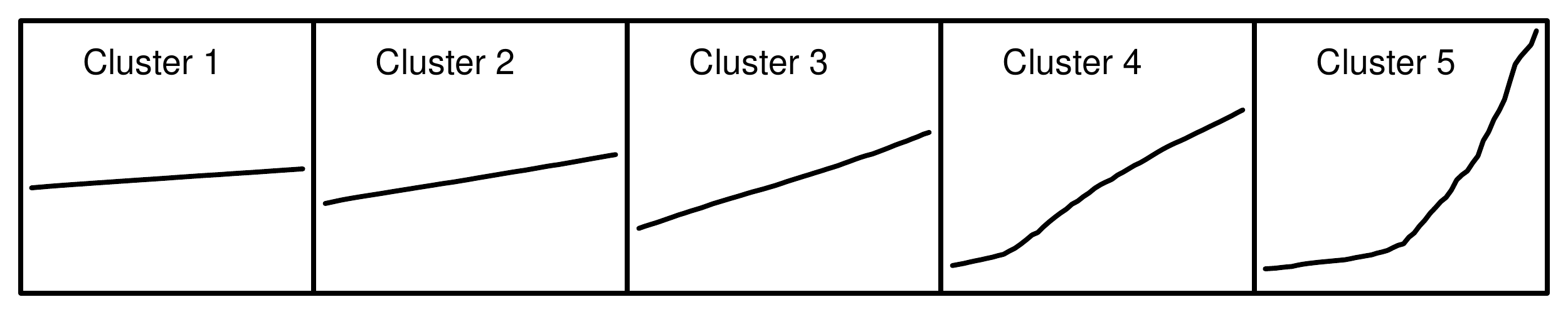}
  \caption{Five growth trends (clusters) identified for the repositories in our dataset}
\label{fig:centroids}
\end{figure}

Figure~\ref{fig:centroids} shows the time series representing the clusters' centroids of the five clusters.
The trends presented by clusters C1, C2, and C3 suggest a linear growth in the number of stars.
The trend presented by cluster C4 differs from the first three ones due variations in the number of stars over the time.
Finally, cluster C5 suggests a viral growth.

As presented in Table~\ref{tab:clusters-description}, cluster C1 concentrates almost half of the repositories in our dataset (49.1\%) while cluster C5 has the lowest concentration (1.2\%).
Table~\ref{tab:clusters-description} also presents the percentage of growth of each cluster, considering the centroids time series.
This percentage ranges from 19.9\% (cluster C1) to 1,659.1\% (cluster C5).

\begin{table}[!h]
  \centering
  \caption{Popularity Trends Description}
\label{tab:clusters-description}
  \begin{tabular}{@{}crrl@{}}
    \toprule
    \multicolumn{1}{c}{\bf Cluster} & \multicolumn{1}{c}{\bf \# Repositories} & \multicolumn{1}{c}{\bf \% Growth}\\
    \midrule
    C1	& 2,087 (49.1\%) & 19.9 \\
    C2  & 1,456 (34.2\%) & 61.3 \\
    C3	& 521  (12.2\%) & 175.1 \\
    C4	& 131  (3.0\%) & 883.2 \\
    C5	& 53   (1.2\%) & 1,659.1 \\
    \bottomrule
  \end{tabular}
\end{table}

When answering RQ \#2, we do not consider specific models for cluster C5 due to two main reasons: (a) it includes only 53 repositories (1.2\%); (b) as presented in Figure~\ref{fig:centroids}, the time series in this cluster do not have a linear shape.\\

\noindent{\bf Repositories Ranking:} To answer RQ \#3, we compute three rankings:
(i) repositories sorted according to the predicted number of stars
using a generic prediction model (configured with $t_{r} = 26$  weeks and $t = 52$); (ii) repositories sorted according to the predicted number of stars
using specific prediction models (configured with $t_{r} = 26$  weeks and $t = 52$); (iii) repositories sorted according
to their real number of stars, as provided by GitHub API, on April 25, 2016, i.e., the last week
we consider to build the time series of stars. In the first two rankings, the rank positions
range from 1 to 4,248 (which is the dataset size). However, the third ranking includes all repositories in the previous rankings plus 468 repositories that entered the list of the most popular repositories in the year before April 25, 2016.
As examples, we have {\sc apple/swift} and {\sc Netflix/falcor}. {\sc Netflix/falcor} is not among the top-5,000 most popular repositories on April 25, 2015, when we
select the repositories used in the study, but it gained popularity to the
point of being the 481st most popular repository one year later. {\sc apple/swift} was created
on October 10, 2015; despite this it is the 23rd most popular repository on
April 25, 2016, when we define the real ranking.
In this way, the investigation conducted to answer RQ \#3 includes the cases where a repository falls in the ranking not only due to a better performance of the repositories used to produce the prediction models, but also due to the performance of any other repository.

\begin{table*}[!h]
  \centering
  \caption{Number of stars gained (real and predicted measures) for the top-10 (first table half) and bottom-10 repositories (second table half). Predictions are produced using a generic model ($t_{r} = 26$  weeks and $t = 52$). We can see that the error (column \aspas{\% Diff}) is lower for the top-repositories.}
\label{tab:generic-example}
  \begin{tabular}{@{}lrrrl@{}}
    \toprule
    \multicolumn{1}{c}{\textbf{Repository}} & \multicolumn{1}{c}{\textbf{Stars}} & \multicolumn{1}{c}{\textbf{Predicted}} & \multicolumn{2}{c}{\textbf{\% Diff}}\\
    \toprule
    {\sc	jquery/jquery		} &	6,160	& 	5,369	& 	-12.84 & \sbar{1284}{10000}	\\
    {\sc	robbyrussell/oh-my-zsh		} &	13,536	& 	11,829	& 	-12.61 & \sbar{1182}{10000}	\\
    {\sc	airbnb/javascript		} &	17,026	& 	14,882	& 	-12.59 & \sbar{1259}{10000}	\\
    {\sc	h5bp/html5-boilerplate		} &	4,896	& 	4,691	& 	-4.19 & \sbar{419}{10000}	\\
    {\sc	meteor/meteor		} &	9,919	& 	10,082	& 	+1.64 & \sbar{164}{10000}	\\
    {\sc	torvalds/linux		} &	10,566	& 	9,682	& 	-8.37 & \sbar{837}{10000}	\\
    {\sc	daneden/animate.css		} &	10,492	& 	9,452	& 	-9.91 & \sbar{991}{10000}	\\
    {\sc	facebook/react-native		} &	18,443	& 	19,373	& 	+5.04 & \sbar{504}{10000}	\\
    {\sc	rails/rails		} &	5,701	& 	5,128	& 	-10.05 & \sbar{1005}{10000}	\\
    {\sc	docker/docker		} &	10,268	& 	9,721	& 	-5.33 & \sbar{533}{10000}	\\
    \midrule
    {\sc	ReactiveRaven/jqBootstrapValidation		} &	213	& 	298	& 	+39.91	& \sbar{3991}{10000} \\
    {\sc	infinitered/ProMotion		} &	119	& 	238	& 	+100.00 & \sbar{10000}{10000}	\\
    {\sc	nslocum/design-patterns-in-ruby		} &	640	& 	731	& 	+14.22 & \sbar{1422}{10000}	\\
    {\sc	jbt/markdown-editor		} &	621	& 	744	& 	+19.81 & \sbar{1981}{10000}	\\
    {\sc	mumble-voip/mumble		} &	565	& 	667	& 	+18.05 & \sbar{1805}{10000}	\\
    {\sc	Manabu-GT/ExpandableTextView		} &	676	& 	623	& 	-7.84 & \sbar{784}{10000}	\\
    {\sc	apache/flink		} &	890	& 	712	& 	-20.00 & \sbar{2000}{10000}	\\
    {\sc	mafintosh/mongojs		} &	322	& 	381	& 	+18.32 & \sbar{1832}{10000}	\\
    {\sc	rofl0r/proxychains-ng		} &	813	& 	790	& 	-2.83 & \sbar{283}{10000}	\\
    {\sc	mikeflynn/egg.js		} &	584	& 	793	& 	+35.79 & \sbar{3579}{10000}	\\
    \bottomrule
  \end{tabular}
\end{table*}

\section{Results}
\label{sec:results}

\vspace{1em} \noindent \emph{\textbf{RQ \#1}: What is the accuracy of the generic prediction models?} \\

\begin{figure}[!h]
  \centering
  \includegraphics[width=0.65\columnwidth, trim={0 2em 0 5em}, clip]{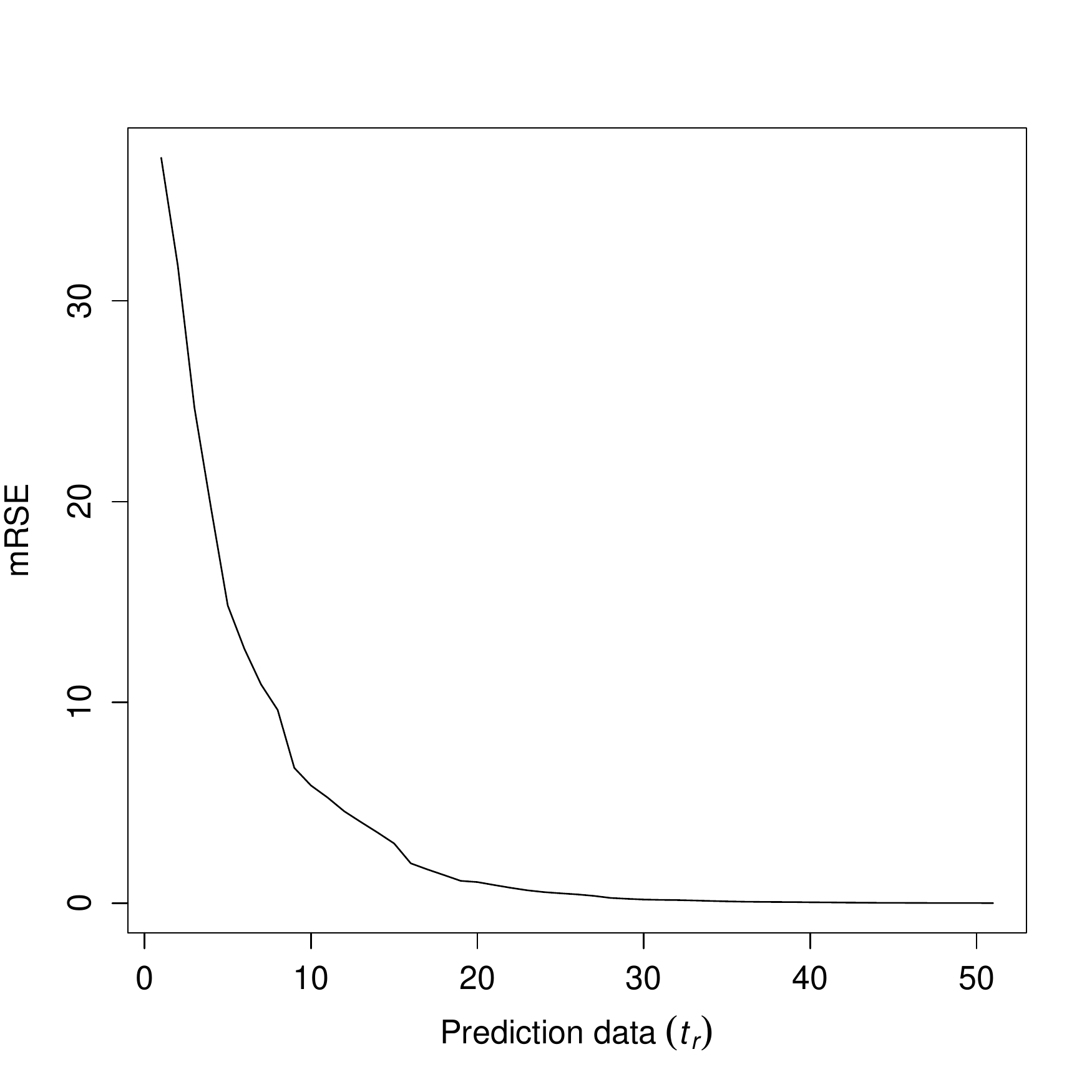}
  \caption{Generic model error}
\label{fig:generic-mrse}
\end{figure}

In order to start answering this question, we produce {\em generic} prediction models and assess their accuracy using 10-fold cross validation for different values of $t_{r}$ (prediction data, see Figure~\ref{fig:cross-validation}).
In all cases, we use the models to predict the number of stars at week 52 ($t = 52$, in Figure~\ref{fig:cross-validation}).
In other words, we use the number of stars in the first $t_{r}$ weeks to predict the number of stars in the 52nd week (last week we considered when collecting the number of stars).
Figure~\ref{fig:generic-mrse} reports the average error (mRSE) across all models.

For small values of $t_{r}$ the models do not perform well, e.g., for $t_{r} = 10$ weeks mRSE $= 5.858 \pm 4.372$ (mean $\pm$ 95\% confidence interval).
However, as we increase the values of $t_{r}$, the results are more accurate.
For example, mRSE $= 0.432 \pm 0.257$ for $t_{r} = 26$ weeks.
This means that we can predict with a low error the number of stars six months ahead, using as training data the past six months of stars.

\begin{figure}[!h]
  \centering
  \includegraphics[width=0.65\columnwidth, trim={0 2em 0 5em}, clip]{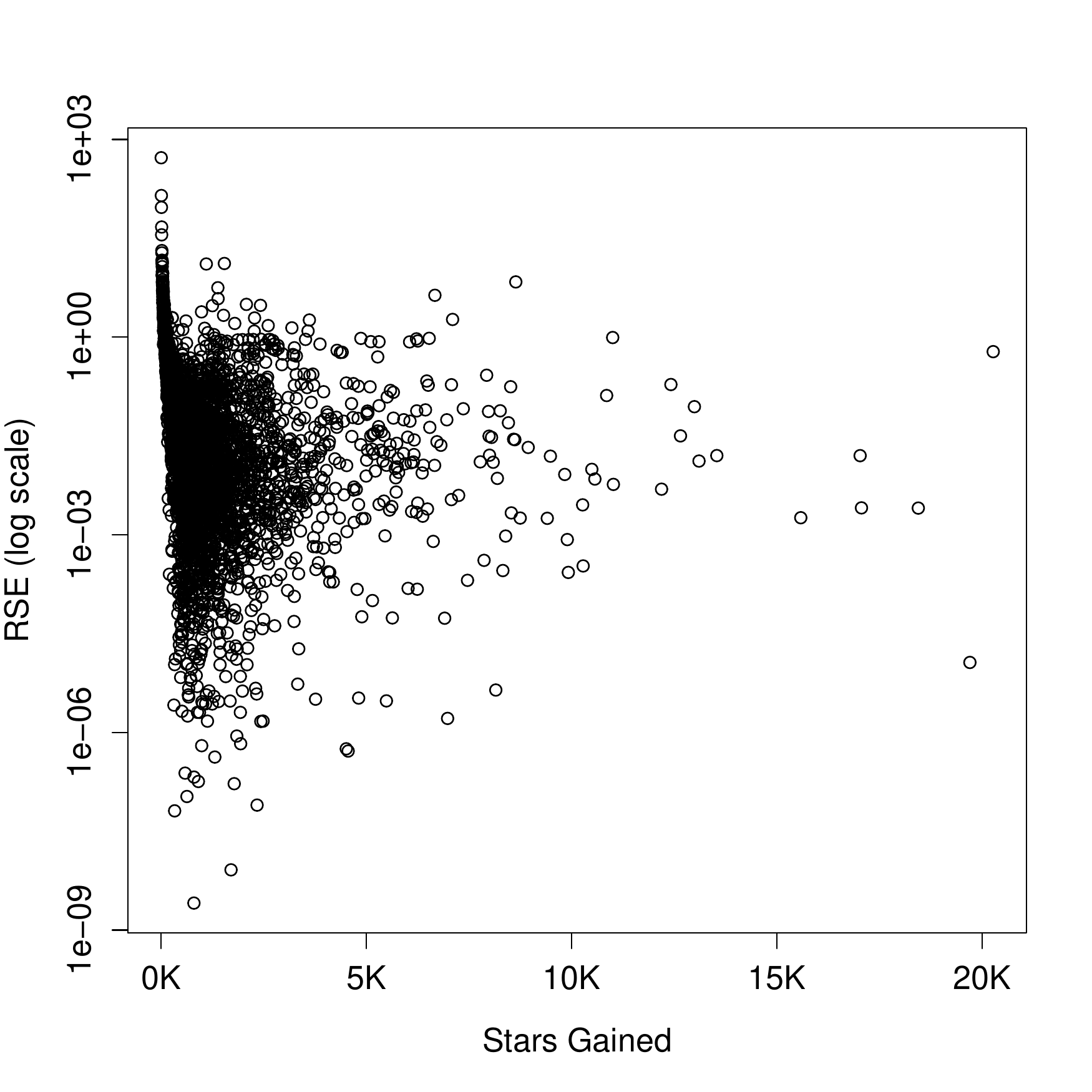}
  \caption{Stars  vs RSE (generic model, $t_{r} = 26$ weeks)}
\label{fig:generic-correlation}
\end{figure}

Figure~\ref{fig:generic-correlation} shows a scatter plot that correlates the number of stars gained and the RSE for the generic models produced using $t_{r} = 26$ weeks.
Each point in this figure represents a repository.
We ran Spearman's rank correlation test and the resulting correlation coefficient $\mathit{rho}$ is -0.50, with $\emph{p-value} < 0.001$. Therefore, the generic models are more accurate for the repositories that gained many stars in the period.

Table~\ref{tab:generic-example} lists the prediction results for the top-10 and bottom-10 repositories with more stars in our dataset.
The column \aspas{Stars} shows the real number of stars gained in 52 weeks and the column \aspas{Predicted} presents the number of stars predicted for the same period using a generic model ($t_{r} = 26$ weeks and $t = 52$).
The difference between the real and the predicted values ranges from 1.64\% to 12.84\%, in absolute values, for the top-10 repositories and from 2.83\% to 100\% for the bottom-10 ones.

\begin{figure*}[!ht]
  \centering
  \begin{subfigure}[b]{0.25\linewidth}
  \includegraphics[width=1\linewidth, trim={0em 2em 2em 5em}, clip, page=1]{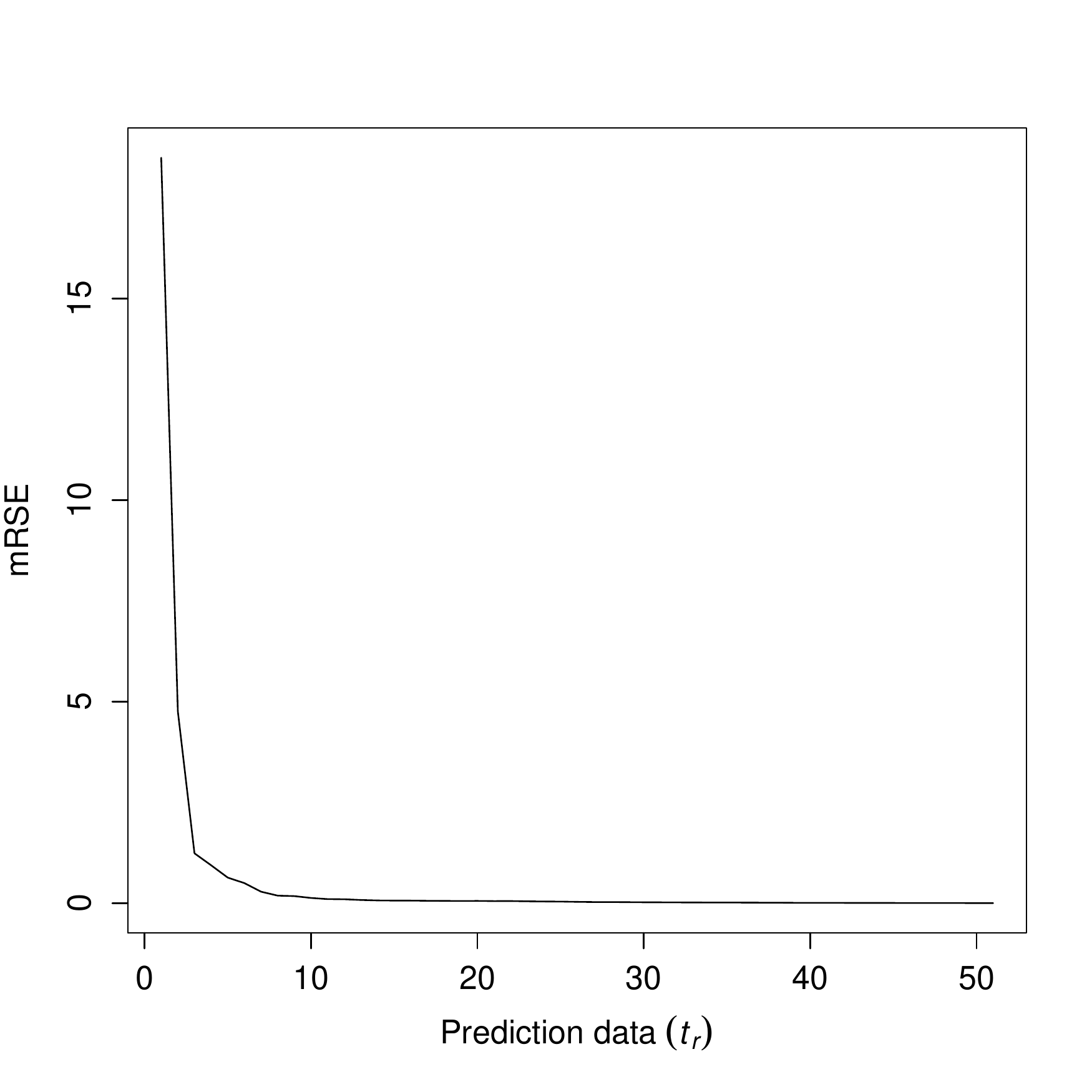}
  \caption{Cluster C1}
  \end{subfigure}%
  \begin{subfigure}[b]{0.25\linewidth}
  \includegraphics[width=1\linewidth, trim={0em 2em 2em 5em}, clip, page=2]{images/mRSE_52weeks_k5_ordered.pdf}
  \caption{Cluster C2}
  \end{subfigure}%
  \begin{subfigure}[b]{0.25\linewidth}
  \includegraphics[width=1\linewidth, trim={0em 2em 2em 5em}, clip, page=3]{images/mRSE_52weeks_k5_ordered.pdf}
  \caption{Cluster C3}
  \end{subfigure}%
  \begin{subfigure}[b]{0.25\linewidth}
  \includegraphics[width=1\linewidth, trim={0em 2em 2em 5em}, clip, page=4]{images/mRSE_52weeks_k5_ordered.pdf}
  \caption{Cluster C4}
  \end{subfigure}%
  \caption{Model prediction error for different growth trends (i.e., clusters extracted using the KSC algorithm)}
\label{fig:specific-mrse}
\end{figure*}

\begin{figure}[!h]
  \centering
  \includegraphics[width=0.65\columnwidth, trim={0 2em 0 5em}, clip]{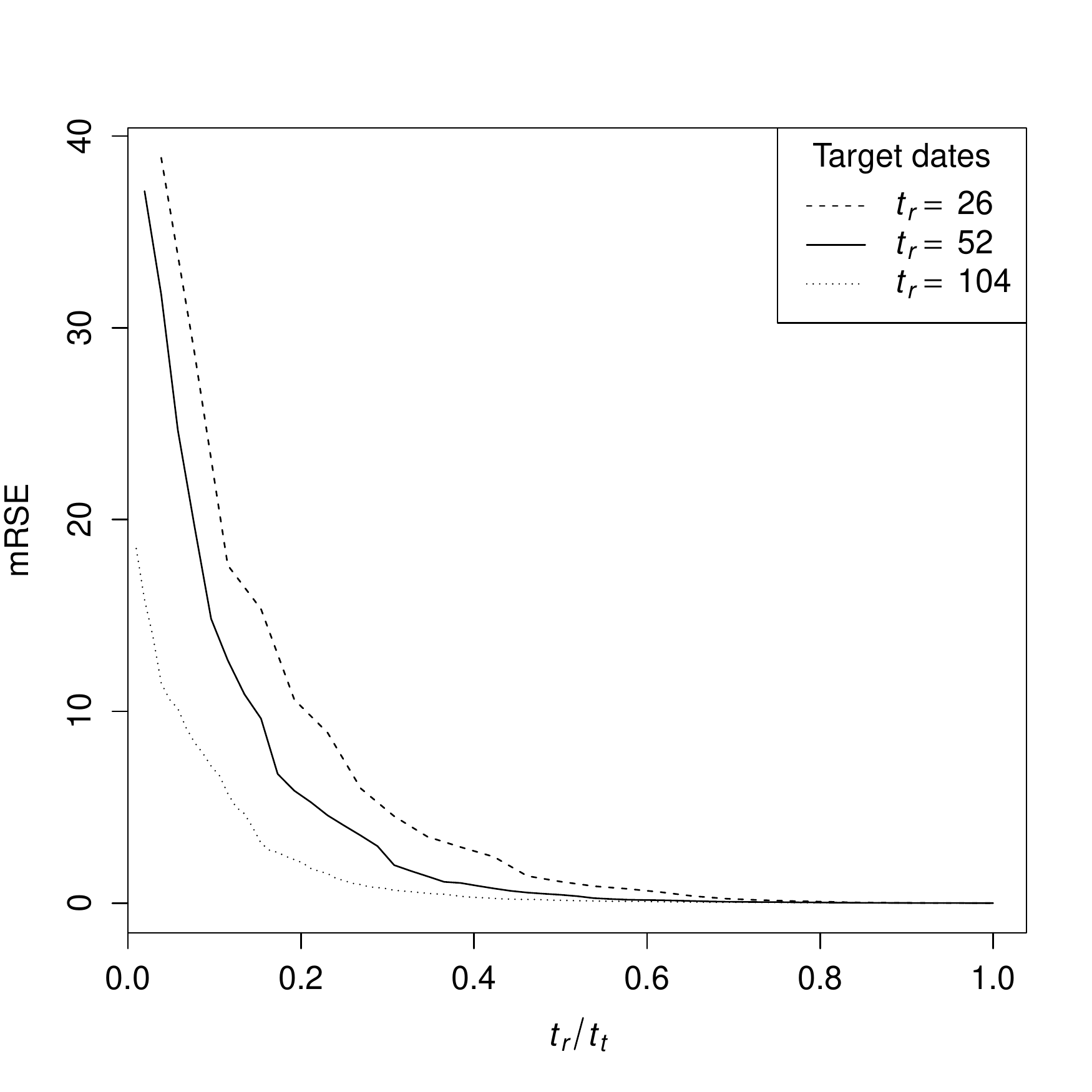}
  \caption{Generic model error (y-axis). Predictions for 26, 52, and 104 weeks, using different fractions of data (x-axis)}
\label{fig:generic-mrse2}
\end{figure}

Finally, we evaluate the accuracy of the generic prediction models for different values of the target week $t$.
Figure~\ref{fig:generic-mrse2} shows the average error for $t = 26$ weeks (half year) and $t = 104$ weeks (two years).
The figure also includes the average error for $t = 52$ weeks (already presented in Figure~\ref{fig:generic-mrse}).
In all cases, we see a decreasing trend of the average error measure.
However, for higher values of $t$, we need less prediction data to achieve similar average errors.
For example, using $t = 52$ weeks and a fraction of time equals to 0.5 (i.e, 26 weeks) the average error (mRSE) is $0.432 \pm 0.257$.
For $t = 104$ weeks, a similar average error (mRSE $= 0.460 \pm 0.182$) happens for a fraction of time of 0.36 (i.e., 38 weeks).\\

\begin{center}
\noindent \fbox{\parbox{0.46\textwidth}{
{\em Summary: } The generic models start to provide accurate predictions when they are trained with data from six months and used to predict the number of stars six months ahead. Furthermore, generic models for highly popular repositories are more accurate than the ones generated for repositories with few stars. }}
\end{center}

\begin{table*}[!th]
  \centering
  \caption{Number of stars gained (real and predicted measures) for the top-10 (first table half) and bottom-10 repositories (second table half). Predictions are produced using specific models ($t_{r} = 26$  weeks and $t =$ week $52$). Column \aspas{\% Improve} shows the gains achieved by specific models, when compared with the predictions provided by generic models. Black bars represent positive gains and gray bars denote negative gains.}
\label{tab:specific-example}
  \begin{tabular}{@{}lcrrrrl@{}}
    \toprule
    \multicolumn{1}{c}{\textbf{Repository}} & \multicolumn{1}{c}{\textbf{Cluster}}  & \multicolumn{1}{c}{\textbf{Stars}} & \multicolumn{1}{c}{\textbf{Predicted}} & \multicolumn{1}{c}{\textbf{\% Diff}} & \multicolumn{2}{c}{\textbf{\% Improve}} \\
    \toprule
    {\sc	jquery/jquery		} & C1 &	6,160	& 	5,578	& 	-9.45 &  +3.39	& \sbar{339}{8235} \\
    {\sc	robbyrussell/oh-my-zsh		} & C2 &	13,536	& 12,826		& 	-5.25 & +7.37	& \sbar{737}{8235} \\
    {\sc	airbnb/javascript		} & C2 &	17,026	& 20,140		& 	+18.29 & -5.70	 & \sbarn{570}{8235} \\
    {\sc	h5bp/html5-boilerplate		} & C1 &	4,896	& 4,690		& 	-4.21 & -0.02 	& \sbarn{2}{8235} \\
    {\sc	meteor/meteor		} & C2 &	9,919	& 	10,571	& 	+6.57 & -4.93	& \sbarn{493}{8235} \\
    {\sc	torvalds/linux		} & C2 &	10,566	& 10,498		& 	-0.64 & +7.72	& \sbar{772}{8235} \\
    {\sc	daneden/animate.css		} & C2 &	10,492	& 	10,045	& 	-4.26 & +5.65	& \sbar{565}{8235} \\
    {\sc	facebook/react-native		} & C3 &	18,443	& 	18,432	& 	-0.06 & +4.98	& \sbar{498}{8235} \\
    {\sc	rails/rails		} & C1 &	5,701	& 	5,386	& 	-5.53 & +4.53	& \sbar{453}{8235} \\
    {\sc	docker/docker		} & C2 &	10,268	& 9,468		& 	-7.79 & -2.46	& \sbar{246}{8235} \\
    \midrule
    {\sc	ReactiveRaven/jqBootstrapValidation		} & C1 &	213	& 	209	& 	-1.88 & +38.03 & \sbar{3803}{8235} \\
    {\sc	infinitered/ProMotion		} & C1 &	119	& 	155	& 	+30.25 & +69.75	& \sbar{6975}{8235} \\
    {\sc	nslocum/design-patterns-in-ruby		} & C3 &	640	& 922	& +44.06 & -29.84	& \sbarn{2984}{8235} \\
    {\sc	jbt/markdown-editor		} & C2 &	621	& 	667	& +7.41 & +12.40 	& \sbar{1240}{8235} \\
    {\sc	mumble-voip/mumble		} & C2 &	565	& 583		& +3.19 & +82.35	& \sbar{8235}{8235} \\
    {\sc	Manabu-GT/ExpandableTextView		} & C3 &	676	& 729		& +7.84 & 0	& \sbar{0}{8235} \\
    {\sc	apache/flink		} & C3 &	890	& 853	& -4.04 & +14.29	& \sbar{1429}{8235} \\
    {\sc	mafintosh/mongojs		} & C1 &	322	& 309	& -4.04 & +14.29	& \sbar{1429}{8235} \\
    {\sc	rofl0r/proxychains-ng		} & C3 &	813	& 886	& +8.98 & -6.15	& \sbarn{615}{8235} \\
    {\sc	mikeflynn/egg.js		} & C1 &	584	& 523	& -10.45 & +25.34	& \sbar{2534}{8235} \\
    \bottomrule
  \end{tabular}
\end{table*}

\vspace{1em}\noindent \emph{\textbf{RQ \#2}: What is the accuracy of the specific prediction models?} \\
\label{sec:results:rq2}

In this second research question, we generate {\em specific} prediction models for the repositories in each cluster (presented in Section~\ref{sec:study-design}) and assess their accuracy using 10-fold cross validation for different values of $t_{r}$.
As in the first question, we predict the number of stars at week 52.

Figure~\ref{fig:specific-mrse} reports the average error across all specific models.
Cluster C1, which concentrates almost half of the repositories, presents a fast decreasing in the average error, e.g., mRSE $= 18.500 \pm 14.501$ for $t_{r} = 1$ week and $0.127 \pm 0.020$ for $t_{r} = 10$ weeks.
This suggests that specific models for this cluster require very few data to provide accurate predictions.
Cluster C2 also presents accurate results for any value of $t_{r}$.
As we can observe, the accuracy of the models for cluster C2 is better than the accuracy for C1 when considering $t_{r} \leq 6$ weeks.
However, for $t_{r} > 6$ weeks, the accuracy of C2 is slightly lower.
For example, mRSE $= 0.030 \pm 0.009$ ($t_{r} = 26$ weeks) and mRSE $= 0.038 \pm 0.009$ ($t_{r} = 26$ weeks) for clusters C1 and C2, respectively.
Cluster C3, which presents the fastest linear trend, shows an initial increasing in the average error, followed by a drastic reduction.
This happens due to inaccurate results of two repositories: {\sc tessalt/echo-chamber-js} (RSE = $284.29$) and {\sc gilesbowkett/rewind} (RSE = $224.48$), which gained a high number of stars at weeks 8 and 13, respectively.\footnote{Because cluster C5 does not follow a linear trend it is not included in our analysis.}

\begin{figure}[!h]
  \centering
  \includegraphics[width=0.65\columnwidth, trim={0em 2em 0em 5em}, clip]{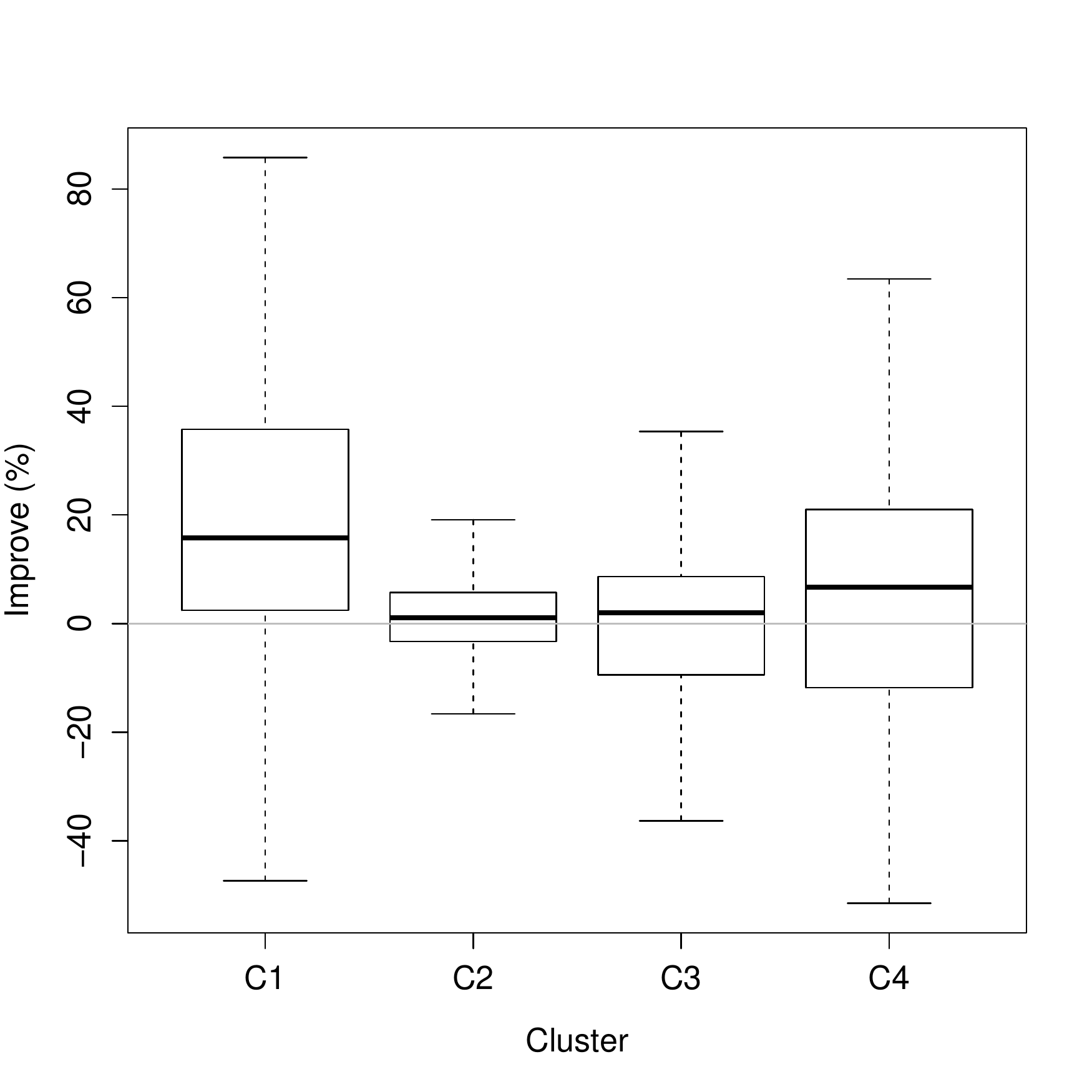}
  \caption{Improvement of specific models per cluster (outliers are omitted).}
\label{fig:specific-improvement-clusters}
\end{figure}

Figure~\ref{fig:specific-improvement-clusters} shows boxplots with the improvements per cluster.
The improvements are calculated from the gains achieved by specific models ($t_{r} = 26$ weeks).
As we can observe, specific models improve the predictions in all clusters, considering the median values.
The median improvements for each cluster are 15.72\%, 1.08\%, 2.00\%, and 6.66\%, respectively.
The repositories in cluster C1 take more advantage of specific models (1st quartile = 2.43\%).
By contrast, clusters C3 and C4 have the highest percentage of repositories with a worst performance (1st quartile equal to -9.46\% and -11.79\%, respectively).


Table~\ref{tab:specific-example} lists the specific prediction results for the top-10 and bottom-10 repositories with more stars in our dataset.
The column \aspas{Stars} shows the real number of stars gained in 52 weeks and the column \aspas{Predicted} presents the number of stars predicted for the same period using specific models ($t_{r} = 26$ weeks).
The difference between the real and predicted number of stars ranges from 0.06\% to 18.29\%, in absolute values, for the top-10 repositories.
Column \aspas{\% Improve} shows the gains achieved by specific models, when compared with the predictions provided by the generic models.
As we can see, the specific models increase the accuracy of the predictions for six out of ten repositories, from 3.39\% to 7.72\%.
For the bottom-10 repositories, the difference between the real and predicted values ranges from 1.88\% to 44.06\% (column \aspas{\% Diff}).
In this case, the specific models produced more accurate results for seven out of ten repositories (column \aspas{\% Improve}).

\begin{center}
\noindent \fbox{\parbox{0.46\textwidth}{
{\em Summary: } Repositories in cluster C1 (slow growth, 49.1\% of the repositories) demand less data to produce reliable predictions. Cluster C1 has also the highest percentage of repositories taking advantage of specific models. Furthermore, the specific models improved the predictions of six (out of ten) top systems, from 3.39\% to 7.72\%.
They also improved the predictions of seven (out of ten) bottom systems, from 1.88\% to 44.06\%.
Therefore, specific models are recommended for repositories with slow growth (cluster C1) and/or among the ones with less stars.}}
\end{center}

\vspace{1em}\noindent \emph{\textbf{RQ \#3}: What is the accuracy of the repositories rank as predicted using the generic and specific models?} \\

\begin{figure}[!h]
  \begin{subfigure}[b]{0.5\columnwidth}
  \includegraphics[width=1\columnwidth, trim={0em 0em 0em 0em}, clip, page=1]{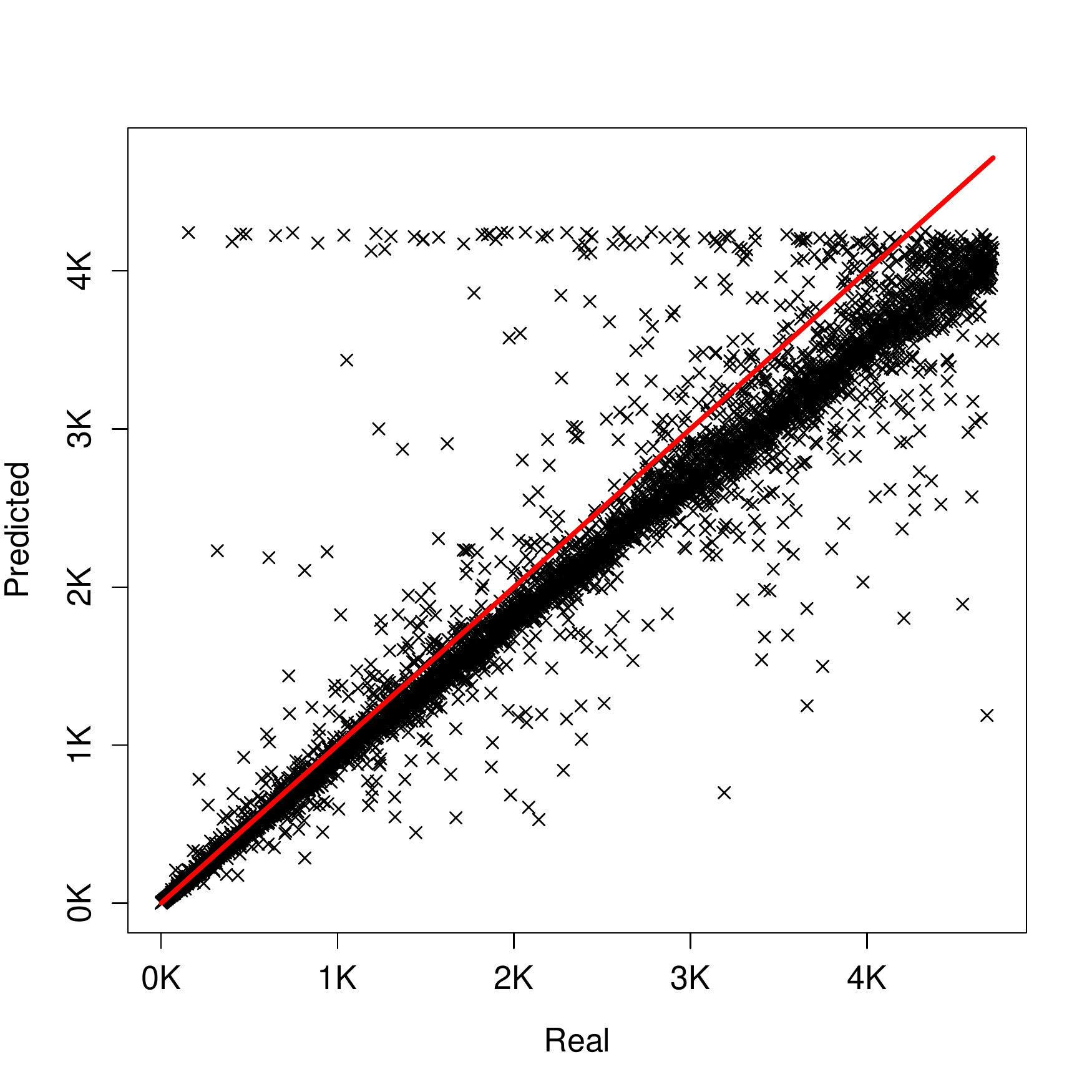}
  \caption{Using generic models}
\label{fig:prediction-points-generic}
  \end{subfigure}%
  \begin{subfigure}[b]{0.5\columnwidth}
  \includegraphics[width=1\columnwidth, trim={0em 0em 0em 0em}, clip, page=2]{images/prediction.pdf}
  \caption{Using specific models}
\label{fig:prediction-points-specific}
  \end{subfigure}%
  \caption{Real vs predicted rankings. The red line is the identity function.}
\label{fig:prediction-points}
\end{figure}

Figure~\ref{fig:prediction-points} shows scatter plots correlating the real rank and predicted rankings using generic and specific models.
The red line represents the identity function, i.e., a perfect match between the real and predicted ranks.
Points above this line are repositories where the predicted rank is higher than the real one (we refer to this kind of error as an underestimation; e.g., a repository is predicted at the 10th position, but in fact it is in position 5th). By contrast, points below the identity line have a predicted rank lower than the real one (we refer to this error as an overestimation; e.g., a repository is predicted at the 5th position, but in fact it in in position 10th).
Initially, we can observe that both models tend to overestimate many predictions, i.e., we usually have more points below the identity line.
This happened because 468 repositories were created and/or quickly became more popular than the ones in our dataset. These repositories are called newcomers, in the context of this research question. Suppose for example that a newcomer appears at rank $i$; in this case it increases the rankings of all repositories with a predicted rank greater than $i$. This shift in the rankings is not detected by the prediction models we investigate, since they do not have information about new systems appearing in the rankings. However, we decide to consider newcomers in this first part of RQ \#3 to simulate a situation that will appear in the practice.



\begin{figure}[!h]
  \centering
  \includegraphics[width=1\columnwidth, trim={0em 0em 0em 0em}, clip]{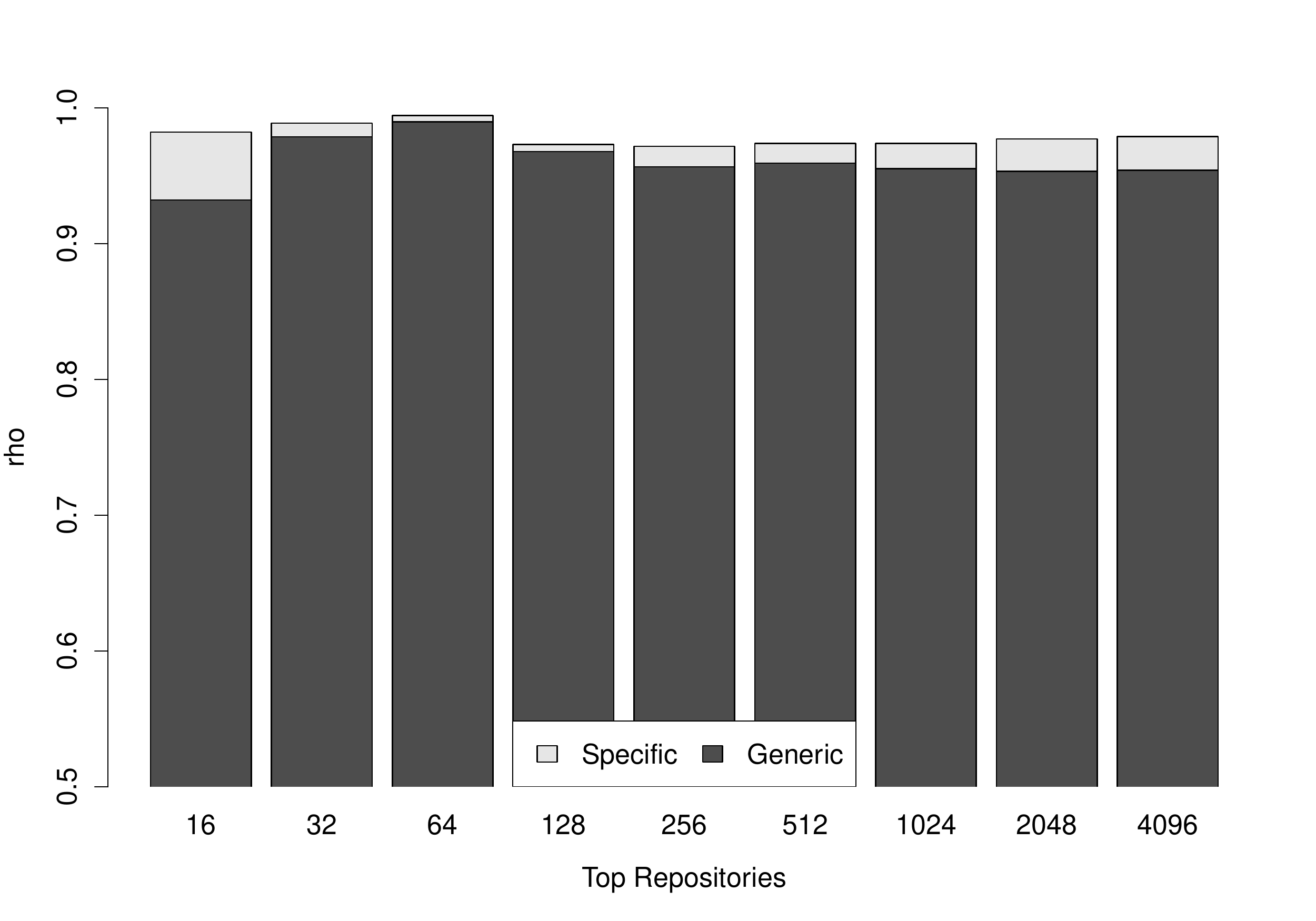}
  \caption{Spearman's rank correlation $\mathit{rho}$ between predicted and real rankings per group of top-repositories (\emph{p-value} $< 0.001$).}
\label{fig:generic-spearman-rho}
\end{figure}

To compare the real and the predicted rankings, we use the Spearman's correlation test.
Since the test requires as input two vectors with the same size, we removed the newcomers from the real rankings.
 For the generic model, we found a strong correlation between the ranks ($\mathit{rho} = 0.9534$ and \emph{p-value} $< 0.001$).
For the specific models, the correlation is slightly better ($\mathit{rho} = 0.9777$ and \emph{p-value} $< 0.001$).
Figure~\ref{fig:generic-spearman-rho} shows the Spearman's coefficient for different groups of top-repositories.
For the top-16 repositories, the correlation using the generic model is lower than the one using the specific models ($\mathit{rho} = 0.9321$ and $\mathit{rho} = 0.9821$, respectively).
For the other top-values values, this difference decreases.
However, the rankings as predicted by the specific models present slightly better results in all cases.

\begin{table*}[!h]
  \centering
  \caption{Real and predicted rankings for the top-10 (first table half) and bottom-10 repositories (second table half), using the generic and the specific models. Marks \aspas{---} indicate repositories that were created and/or became popular after the date we set to select the repositories considered in this study.}
\label{tab:prediction-example}
  \begin{tabular}{@{}lccrrcrr@{}}
    \toprule
    \multicolumn{1}{c}{\multirow{2}{*}{\textbf{Repository}}} & \multicolumn{1}{c}{\multirow{2}{*}{\textbf{Real}}} & & \multicolumn{2}{c}{\textbf{Predicted}} & &  \multicolumn{2}{c}{\multirow{1}{*}{\textbf{Diff}}} \\
    \cmidrule{4-5}\cmidrule{7-8}
    & & & \multicolumn{1}{c}{\textbf{Generic}} & \multicolumn{1}{c}{\textbf{Specific}} & &  \multicolumn{1}{c}{\textbf{Generic}} & \multicolumn{1}{c}{\textbf{Specific}} \\
    \toprule
    {\sc	jquery/jquery	} &	1	& &	1	&	1	& &	0	&	0	\\
    {\sc	robbyrussell/oh-my-zsh	} &	2	& &	2	&	2	& &	0	&	0	\\
    {\sc	airbnb/javascript	} &	3	& &	5	&	3	& &	2	&	0	\\
    {\sc	h5bp/html5-boilerplate	} &	4	& &	4	&	5	& &	0	&	1	\\
    {\sc	meteor/meteor	} &	5	& &	3	&	4	& &	-2	&	-1	\\
    {\sc	torvalds/linux	} &	6	& &	8	&	6	& &	2	&	0	\\
    {\sc	daneden/animate.css	} &	7	& &	9	&	8	& &	2	&	1	\\
    {\sc	facebook/react-native	} &	8	& &	6	&	7	& &	-2	&	-1	\\
    {\sc	rails/rails	} &	9	& &	10	&	10	& &	1	&	1	\\
    {\sc	docker/docker	} &	10	& &	11	&	11	& &	1	&	1	\\
    \midrule
    {\sc	jbt/markdown-editor	} &	4,707	& &	3,908	&	4,001	& &	-799	&	-706	\\
    {\sc	apache/flink	} &	4,708	& &	4,180	&	4,167	& &	-528	&	-541	\\
    {\sc	google/ion	} &	4,709	& &	---	&	---	& &	---	&	---	\\
    {\sc	Manabu-GT/ExpandableTextView	} &	4,710	& &	4,149	&	4,016	& &	-561	&	-694	\\
    {\sc	iPaulPro/Android-ItemTouchHelper-Demo	} &	4,711	& &	---	&	---	& &	---	&	---	\\
    {\sc	mumble-voip/mumble	} &	4,712	& &	4,011	&	4,092	& &	-701	&	-620	\\
    {\sc	mafintosh/mongojs	} &	4,713	& &	4,080	&	4,164	& &	-633	&	-549	\\
    {\sc	mikeflynn/egg.js	} &	4,714	& &	3,569	&	4,194	& &	-1,145	&	-520	\\
    {\sc	wequick/Small	} &	4,715	& &	---	&	---	& &	---	&	---	\\
    {\sc	rofl0r/proxychains-ng	} &	4,716	& &	4,136	&	3,824	& &	-580	&	-892	\\

    \bottomrule
  \end{tabular}
\end{table*}

Table~\ref{tab:prediction-example} shows the predicted rank for the top-10 and bot\-tom-10 repositories in our dataset.
The column \aspas{Real} represents the real rank in GitHub when our dataset was collected.
The column \aspas{Predicted} shows the predicted rank using the generic (column \aspas{Generic}) and the specific models (column \aspas{Specific}).
Repositories that were created or became popular after the date we start collecting the time series are marked with \aspas{---}.
Finally, the column \aspas{Diff} shows the difference between the predicted and the real rankings.
As mentioned, both predictions are more accurate for the top repositories and tend to overrate the rank of the bottom repositories.
For the top-10 repositories, the difference in absolute values between the ranks ranges from 0 to 2 (generic models) and from 0 to 1 (specific models).
For the bottom-10 repositories, the difference between the ranks ranges from 528 to 1,145 (generic models) and from 520 to 892 (specific models).
However, it is important to notice that the distribution of the number of stars per repository is heavy-tailed~\cite{Borges2015}. Therefore, minor differences
in the predicted number of stars can represent a movement of hundreds of positions in the relative order of a repository.
For example, an error of 175 stars in the number of stars predicted to
{\sc Apache/Flink} is responsible to generate a error of 528 positions
in its ranking (from position 4,708 in the real ranking; to position
4,180 in the ranking predicted using a generic model).

\begin{center}
\noindent \fbox{\parbox{0.46\textwidth}{
{\em Summary: } Prediction models tend to overestimate the repositories ranks, specifically due to
the entry of newcomers in the list of popular repositories. However, when newcomers are not considered,
there is a very strong correlation between predicted and real rankings ($\mathit{rho} = 0.9534$ and $0.9777$,  for generic and specific models, respectively.)
}}
\end{center}

\section{Threats to Validity}
\label{sec:validity}

\noindent{\em Measuring popularity using the number of stars.} In the investigation reported in this paper, we measure popularity using the number of stars of the GitHub repositories, as in other studies~\cite{Weber2014, Aggarwal2014}. However, we highlight that developers can star a repository for other reasons, for example, to create bookmarks.\\[-0.25cm]


\noindent{\em Repositories selection.} GitHub has 17,136,765 public repositories, including forks (in June 15, 2016). For this study, we started with the top-5,000 repositories with more stars and after a cleaning step, we analyze 4,248 repositories. However, we stress that our goal is to predict popularity of most starred repositories. For example, 12,462,551 repositories (73\%) have no stars and probably will never receive one in their lifetime. In other words, it is probably easier (and less useful) to make predictions for a dataset with all public repositories in GitHub.\\[-0.25cm]

\noindent{\em Growth trends}. The selection of the number of clusters is a key parameter in clustering algorithms like KSC. To mitigate this threat, we use the $\beta_{CV}$ heuristic~\cite{Menasce2001} to define the best number $k$ of clusters. We also discard cluster C5 from our evaluation of specific models (RQ \#2), since the repositories in this cluster do not follow a linear growth. Notice, however, that cluster C5 includes only 53 repositories (1.2\%).

\section{Related Work}
\label{sec:related_work}

Our work was inspired by the vast literature on defect prediction.
For example, a systematic literature review listed 208 defect prediction studies~\cite{hall12}, which differ regarding the software metrics used for prediction, the modeling technique, the granularity of the independent variable, and the validation technique.  As independent variables, the studies  use source code metrics (size, cohesion, coupling, etc), change metrics, process metrics, code smells instances, etc. The modeling techniques vary with respect to linear regression, logistic regression, naive bayes, neural networks,  etc. In this paper, instead of predicting the future number of defects of a system, we rely on multiple linear regressions to predict the number of stars of GitHub repositories.

In a previous paper, we studied the popularity of GitHub repositories aiming to answer four research questions~\cite{Hudson2016}. We concluded
that the three most common domains on GitHub are web libraries and frameworks, non-web libraries and frameworks, and software tools.
Additionally, we found that repositories owned by organizations are more popular than the ones owned by individuals (RQ \#1). We also found
a strong correlation between stars and forks and a weak correlation both between stars and commits, and   between stars and contributors (RQ \#2). We concluded that repositories have a tendency to receive more stars right after their first release (RQ \#3). We also showed that there is an acceleration in
the number of stars gained after releases (RQ \#4). In a previous note, we started an investigation about the popularity of GitHub repositories~\cite{Borges2015}. We showed that the number of stars follows a highly skewed distribution. Jian et al. explore {\em why} and {\em how} developers fork {\em what}
from {\em whom} in GitHub~\cite{jiang2016}. They report that  some repository owners are popular,
and attract many forks;  other owners are unpopular and rarely attract forks.
They also present that attractive owners have higher percentage of organizations,
more followers and earlier registration in GitHub. Since forks are relevant
operations in GitHub, future work may investigate prediction models for number of forks.

Martin et al.~\cite{Martin2016FSE} record time-series information about popular Google Play apps  and investigate how release frequency can affect an app's performance, as measured by rating, popularity and number of user reviews. They label as ``impactful releases'' the ones that caused a significant change on the app's popularity, as inferred by Causal Impact Analysis (a form of causal inference). They report that more mentions of features and fewer mentions of bug fixing increase the chance for a release to be impactful. Couto et al.~\cite{jss2014} follow a similar approach but to identify causal relationships between changes in internal measures of software quality (coupling, cohesion, complexity, etc) and the number of defects reported for a system.

Several other studies examine the relationship between popularity of mobile apps and their code properties~\cite{Datta2013, Fu2013, LinaresVasquez2013, Ruiz2014, Tian2015, Guerrouj2015, Palomba2015, Corral2015}.
Yuan et al.~investigate 28 factors along eight dimensions to understand how high-rated Android applications are different from low-rated ones~\cite{Tian2015}. Their result shows that external factors, like number of promotional images, are the most influential factors.
Ruiz et al.~examine the relationship between the number of ad libraries and app's user ratings~\cite{Ruiz2014}. They show that there is no relationship between the number of ad libraries in an app and its rating.
Guerrouj et al.~analyse changes of Android API elements between releases and report that high app churn leads to lower user ratings~\cite{Guerrouj2015}.
Linares-V{\'a}squez et al.~investigate how the fault- and change-proneness of Android API elements relate to applications' lack of success~\cite{LinaresVasquez2013}. They state that making heavy use of fault- and change-prone APIs can negatively impact the success of apps.


Popularity prediction in other social networks is the target of several studies.
In Twitter, Ma et al. predict hashtag popularity to identify fast emerging topics attracting collective attention~\cite{ma2013}. Their results reveal that context features (e.g., number of users that tweeted the hashtag) are relatively more effective than content feature (e.g., number of tweets with the hashtag). Tsur and Rappoport used a hybrid approach based on linear regressions to predict the spread of ideas in Twitter and found that a combination of content features with temporal and topological features minimizes prediction error~\cite{tsur2012}.
In YouTube, Szabo and Huberman found a strong linear correlation between the logarithmically transformed popularity of videos at early and later times. Based on this finding, they present a model to predict future popularity~\cite{Szabo2010}.
Pinto et al. propose two prediction models based on multivariate linear regression that incorporate information about historical patterns~\cite{Pinto2013}.
Finally, Roy et al. propose a framework called {\em SocialTransfer} that utilizes knowledge from social streams (e.g., Twitter) to discover sudden popularity bursts in videos. They show that social trends have a ripple effect as they spread from the Twitter domain to the video domain~\cite{Roy2013}.
To our knowledge, we are the first to target popularity prediction---measured by the number of stars---of software projects in the GitHub social coding network.

\section{Conclusion}
\label{sec:conclusion}

In this paper, we use multiple linear regressions to predict the popularity of GitHub repositories.
We found that general models, i.e., models produced using the top GitHub repositories, start to provide accurate predictions when they are trained with data from six months and used to predict the number of stars six months ahead (RQ \#1). We also found that specific models, i.e., models produced using repositories that share the same growth trend, can reduce the average prediction error and produce reliable predictions using less data. For the most common growth trend in our dataset, which includes almost half of the repositories, specific models improved significantly the accuracy of the predictions (RQ \#2). Finally, we report that prediction models tend to overestimate the repositories ranks. However, when newcomers are not considered, there is a very strong correlation between predicted and real rankings (RQ \#3).
As future work, we plan to extend the specific prediction models to consider different programming languages. We plan to investigate different approaches to predict popularity, for example, epidemic models and machine learning models.
We also plan to investigate predictions for other measures provided by GitHub, such as forks, watchers, and contributors.

\section{Acknowledgments}

\noindent This research is supported by FAPEMIG and CNPq.

%
\bibliographystyle{abbrv}
\bibliography{bibfile}  

\end{document}